\documentclass[%
 reprint,
 superscriptaddress,
 amsmath,amssymb,
 aps,
 prb,
]{revtex4-2}

\usepackage{graphicx}
\usepackage{dcolumn}
\usepackage{physics}
\usepackage{siunitx}
\usepackage[version=4]{mhchem}
\usepackage[colorlinks=true,citecolor=blue,linkcolor=blue,urlcolor=blue]{hyperref}

\DeclareSIUnit{\peratom}{/atom}

\usepackage{cleveref}
\crefname{equation}{Eq.}{Eq.}
\crefname{figure}{Fig.}{Fig.}
\crefname{table}{Table}{Table}
\crefname{section}{Sec.}{Sec.}
\crefname{appendix}{Appendix}{Appendix}

\Crefname{equation}{Equation}{Equation}
\Crefname{figure}{Figure}{Figure}
\Crefname{table}{Table}{Table}
\Crefname{section}{Section}{Section}
\Crefname{appendix}{Appendix}{Appendix}

\begin{document}

\title{Search for high-pressure phases of yttrium via a data assimilation approach}

\author{Yuuki Kubo}
 \email{yuuki.kubo@phys.s.u-tokyo.ac.jp}
\affiliation{Department of Physics, The University of Tokyo, 7-3-1 Hongo, Bunkyo-ku, Tokyo 113-0033, Japan}
\author{Takahiro Ishikawa}
\affiliation{Department of Earth and Space Science, The University of Osaka, 1-1 Machikaneyama, Toyonaka, Osaka 560-0043, Japan}
\author{Yuta Tanaka}
\affiliation{AI Innovation Department, ENEOS Holdings, Inc., 8 Chidoricho, Naka-ku, Yokohama, Kanagawa 231-0815, Japan}
\author{Yuki Nakamoto}
\affiliation{KYOKUGEN, Graduate School of Engineering Science, The University of Osaka, 1-3 Machikaneyamacho, Toyonaka, Osaka 560-8531, Japan}
\author{Masafumi Sakata}
\affiliation{Department of Electrical, Electronic and Computer Engineering, Faculty of Engineering, Gifu University, 1-1 Yanagido, Gifu City 501-1193, Japan}
\author{Shinji Tsuneyuki}
\affiliation{Department of Physics, The University of Tokyo, 7-3-1 Hongo, Bunkyo-ku, Tokyo 113-0033, Japan}
\affiliation{Digital Materials Science Research Team, RIKEN Center for Computational Science (R-CCS), 2-1 Hirosawa, Wako, Saitama 351-0198, Japan}

\date{\today}

\begin{abstract}
We investigate the distorted face-centered-cubic (dfcc) phase of yttrium (\ce{Y}) using a data-assimilation-based structure search that combines high-resolution powder x-ray diffraction (XRD) data with machine-learning interatomic potentials.
By exploring supercells containing up to 128 atoms, we identify three low-enthalpy phases: the previously reported $I4_1/a$ structure and two additional structures, $Ibam$ and $R\overline{3}$.
No data-assimilation-derived structure relaxes to the previously proposed $R\overline{3}m$ phase.
Phonon calculations show that $I4_1/a$, $Ibam$, and $R\overline{3}$ are dynamically stable, whereas $R\overline{3}m$ exhibits imaginary modes near the $\Gamma$ point, indicating dynamical instability.
Enthalpy calculations using both PBE and r$^{2}$SCAN place the four candidate structures within about \SI{10}{\milli\electronvolt}/atom, indicating a complex energy landscape with multiple competing minima, although $R\overline{3}m$ is consistently highest in enthalpy and r$^{2}$SCAN favors $I4_1/a$ throughout the dfcc pressure range.
Rietveld refinements of the powder XRD profile at \SI{60}{\giga\pascal} further narrow the viable structural models to $I4_1/a$ and $Ibam$, both of which reproduce the experimental data better than $R\overline{3}m$ and $R\overline{3}$.
Taken together with the energetic ordering and dynamical stability, these results identify $I4_1/a$ as the most plausible structure of the dfcc phase of \ce{Y}, with $Ibam$ remaining a close competing candidate, particularly toward the high-pressure side of the dfcc region.
\end{abstract}

\maketitle

\section{Introduction}

Yttrium (\ce{Y}) and scandium (\ce{Sc}) are trivalent transition metals that are commonly grouped with the lanthanides as rare-earth metals.
Except for \ce{Ce}, \ce{Eu}, and \ce{Yb}, the trivalent lanthanides undergo a characteristic sequence of pressure-induced structural transitions, namely hexagonal close-packed (hcp) $\rightarrow$ $\alpha$-\ce{Sm}--type $\rightarrow$ double hcp (dhcp) $\rightarrow$ distorted face-centered cubic (dfcc), with increasing pressure~\cite{PhysRevB.11.2836, BENEDICT198614, PhysRevLett.38.564}.
A similar hcp $\rightarrow$ $\alpha$-\ce{Sm}--type $\rightarrow$ dhcp sequence is also observed at ambient pressure across the lanthanide series toward decreasing atomic number.
Because \ce{Y}, despite lacking $4f$ electrons, follows essentially the same structural sequence under compression, these transitions are generally understood to originate primarily from pressure-induced $s$--$d$ electron transfer, i.e., increased occupation of the $d$ band with pressure~\cite{PhysRevLett.38.564, PhysRevLett.47.1065, GROSSHANS1982282, PhysRevB.102.094104}.

At ambient conditions, \ce{Y} crystallizes in the hcp structure.
Upon compression at room temperature, it transforms sequentially into the $\alpha$-\ce{Sm}--type phase near \SI{15}{\giga\pascal}, the dhcp phase near \SI{25}{\giga\pascal}, and the dfcc phase around \SI{50}{\giga\pascal}~\cite{PhysRevLett.47.1065, GROSSHANS1982282, Samudrala_2012, PhysRevB.102.094104}.
Among rare-earth metals, the dfcc phase has often been associated with the rhombohedral $R\overline{3}m$ structure~\cite{NHamaya_1993, PhysRevB.80.134105, SAMUDRALA2013275}, and the same assignment has conventionally been adopted for \ce{Y}~\cite{PhysRevB.84.132101, Samudrala_2012}.
At higher pressures, around \SI{100}{\giga\pascal}, \ce{Y} undergoes a further transformation to another phase that has recently been identified as orthorhombic $Fddd$~\cite{PhysRevB.102.094104, PhysRevB.102.104508}.

On the theoretical side, DFT-based structure searches combined with evolutionary algorithms have clarified several aspects of the high-pressure phase diagram of \ce{Y}~\cite{PhysRevB.84.132101, PhysRevLett.109.157004, Ishikawa02012015, LI2019428}.
For example, Chen \textit{et al.}\ predicted the stability of the $Fddd$ phase above \SI{100}{\giga\pascal} prior to its experimental observation~\cite{PhysRevLett.109.157004, Samudrala_2012, PhysRevB.102.104508}.
Ishikawa \textit{et al.}\ investigated the dfcc region and proposed two candidate structures, $I4_1/a$ and $P\overline{1}$, both of which have slightly lower enthalpies than the conventional $R\overline{3}m$ model~\cite{Ishikawa02012015}.
More recently, calculations using the regularized-restored strongly constrained and appropriately normed (r$^{2}$SCAN) meta-generalized gradient approximation (GGA) functional~\cite{Furness2020}, rather than the standard Perdew--Burke--Ernzerhof (PBE) GGA~\cite{PhysRevLett.77.3865}, were shown to reproduce the transition pressures of the lower-pressure hcp $\rightarrow$ $\alpha$-\ce{Sm}--type $\rightarrow$ dhcp sequence more accurately~\cite{jp7h-d7q3}.
This result suggests that the choice of exchange-correlation functional may also be important for the stability ordering of dfcc-related structures.

Despite these advances, the detailed crystal structure of the dfcc phase of \ce{Y} remains unresolved.
Pace \textit{et al.}\ performed Rietveld refinements of high-pressure powder x-ray diffraction (XRD) data using both the $R\overline{3}m$ and $I4_1/a$ models, but were unable to distinguish the two structures unambiguously~\cite{PhysRevB.102.094104}.
This difficulty reflects a general limitation of high-pressure powder diffraction, where peak overlap, broadening, and preferred orientation can obscure subtle symmetry lowering.
At the same time, previous theoretical searches in the dfcc region were restricted to unit cells containing at most eight atoms, leaving open the possibility that overlooked low-enthalpy structures with larger periodicities may exist.
Resolving this issue is also important from the viewpoint of superconductivity.
Under compression, the superconducting transition temperature $T_\mathrm{c}$ of \ce{Y} increases steadily and reaches \SI{19.5}{\kelvin} at \SI{115}{\giga\pascal}, which is the second highest among the rare-earth elements after \ce{Sc}~\cite{HAMLIN200782, PhysRevB.99.085137, PhysRevB.102.104508, Cao2023, PhysRevLett.130.256002, Matsuoka_2026}.
A reliable structural model for the dfcc phase is therefore essential for understanding the microscopic origin of this pronounced enhancement of superconductivity.

In this study, we combine newly obtained high-resolution powder XRD data with machine-learning interatomic potentials (MLIPs) in a data-assimilation-based structure search, extending the accessible search space to supercells containing up to 128 atoms.
We then perform DFT relaxations, enthalpy comparisons using both PBE and r$^{2}$SCAN, and Rietveld refinements of the experimental diffraction profile to assess the resulting candidate structures.
This approach enables a systematic reexamination of the conventional structural assignment of the dfcc phase of \ce{Y} and provides a general framework for identifying closely competing high-pressure structures from limited powder-diffraction information.
\section{Methods}
\subsection{High-pressure powder XRD measurements}

High-pressure powder XRD measurements were carried out on yttrium using a Mao-type symmetric diamond-anvil cell (DAC).
A yttrium ingot with a purity of 99.9\% was purchased from Nilaco Co.
All sample preparation and loading procedures were performed in an argon atmosphere to minimize oxidation and other unwanted chemical reactions.
Diamond anvils with \SI{100}{\micro\meter} culets were used.
A Re gasket was preindented from an initial thickness of \SI{250}{\micro\meter} to approximately \SI{15}{\micro\meter}, and a sample chamber of \SI{40}{\micro\meter} diameter was drilled at its center.
A piece of yttrium with a maximum lateral size of approximately \SI{15}{\micro\meter} was placed in the sample chamber.
Neon was loaded as the pressure-transmitting medium.
Pressure was determined from the equation of state of Pt~\cite{10.1063/1.344177}.

Angle-dispersive XRD experiments were performed at beamline 16-ID-B of HPCAT, Advanced Photon Source, Argonne National Laboratory.
Monochromatic x-rays with a wavelength of \SI{0.3977}{\angstrom} were used with a \SI{5}{\micro\meter} collimator.
To improve crystallinity and obtain clean diffraction profiles, laser heating was used only for annealing at high pressure, whereas all diffraction data analyzed in the present work were collected at room temperature.
Two-dimensional diffraction images were recorded on a MAR345 image plate (pixel size $100\times 100\,\SI{}{\micro\meter}^2$) with exposure times of 180--430\,\SI{}{\second}.
The diffraction rings were integrated using \textsc{FIT2D}~\cite{Hammersley01011996} to obtain one-dimensional powder diffraction profiles as a function of $2\theta$.
The profile collected at \SI{60}{\giga\pascal} was used in the data-assimilation calculations and in the Rietveld refinements.

\subsection{High-pressure electrical resistance measurements}

Electrical resistance measurements on yttrium were carried out in separate experiments.
A Cu--Be diamond-anvil cell (DAC) equipped with diamond anvils having beveled culets of 50--300~\si{\micro\meter} was used, without a pressure-transmitting medium.
Four-probe electrical contacts were formed at the interface between the sample and the insulating layer using platinum-foil electrodes.
The insulating layer was prepared on a rhenium gasket from \ce{Al2O3} powder without binder in order to minimize possible chemical reactions with the sample.
Electrical resistance was measured using an AC resistance bridge (LR-700) with an excitation current of \SI{3}{\micro\ampere}.
Pressure was determined from the Raman edge of the diamond anvils~\cite{10.1063/1.1778482}.

\subsection{Data-assimilation-based structure exploration and first-principles calculations}

Previous theoretical studies of the dfcc phase of \ce{Y} have largely relied on structure searches based on evolutionary algorithms combined with DFT calculations.
Although powerful, such approaches are computationally expensive and are therefore typically limited to unit cells containing only a few tens of atoms~\cite{PhysRevLett.109.157004, Ishikawa02012015, LI2019428}.
MLIPs have recently become a practical way to accelerate the evaluation of energies and forces in crystal-structure searches~\cite{PhysRevB.99.064114, PhysRevB.106.014102, PhysRevB.109.094106, Ishikawa_2026}.
Their reliability, however, depends critically on both the intrinsic accuracy of the potential and the coverage of the training data.
When experimental data are available, it is advantageous to bias structural exploration toward regions of configuration space that are consistent with experiment~\cite{Putz:zm0055, LANNING2000296, Coelho:ks0007, Meredig2013, GAO201740, PhysRevMaterials.1.063802, ADACHI201992, Lee2023}.

On this basis, we employed a data-assimilation-based structure-search scheme~\cite{PhysRevMaterials.2.053801, 10.1063/5.0125553, ZHAO2023122028, Zarrouk2024, 10.1063/5.0223390, D4DD00269E, zarrouk2025molecularaugmenteddynamicsgenerating}, in which the cost function contains a penalty term that quantifies the mismatch between experimental and simulated diffraction profiles.
The cost function $F$ is defined as
\begin{equation}
F(\mathbf{R})
= E(\mathbf{R})
+ \alpha N\, D\!\left[
I_{\mathrm{exp}}(2\theta),
I_{\mathrm{calc}}(2\theta;\mathbf{R})
\right],
\label{eq:cost}
\end{equation}
where $\mathbf{R}$ denotes the atomic coordinates, $E(\mathbf{R})$ is the interatomic-potential energy, and $D$ is a penalty term that measures the dissimilarity between the experimental diffraction intensity $I_{\mathrm{exp}}$ and the simulated intensity $I_{\mathrm{calc}}$ for configuration $\mathbf{R}$.
Here, $N$ is the number of atoms and $\alpha$ controls the relative weight of energetic stability and agreement with the diffraction data.

Two MLIPs were employed to evaluate $E(\mathbf{R})$ during the search stage.
The first was the PreFerred Potential (PFP)~\cite{10.1038/s41467-022-30687-9}, as implemented in the \textsc{Matlantis} atomic simulator~\cite{Matlantis} (version 6.0.0).
PFP is a universal graph-neural-network potential trained on approximately $6\times10^{7}$ DFT data points spanning 96 elements, including diverse high-pressure environments.
The second was a quadratic Spectral Neighbor Analysis Potential (qSNAP)~\cite{THOMPSON2015316, 10.1063/1.5017641} constructed specifically for the high-pressure dfcc region of \ce{Y}.
The qSNAP model was trained on approximately 10,000 DFT-labeled structures using \textsc{FitSNAP}~\cite{Rohskopf2023}.
The training configurations were sampled from \textit{ab initio} molecular-dynamics (AIMD) trajectories of 64-atom supercells derived from the $I4_1/a$ and $P\overline{1}$ structures~\cite{Ishikawa02012015}.
The AIMD simulations were performed with the projector augmented-wave (PAW) method~\cite{PhysRevB.50.17953, PhysRevB.59.1758} implemented in \textsc{VASP}~\cite{KRESSE199615}, using the PBE GGA, a plane-wave cutoff of \SI{400}{\electronvolt}, and a $2\times2\times2$ $k$-point mesh.
Additional details of the qSNAP training set, fitting parameters, and validation are provided in the Supplemental Material~\cite{SM}.
Independent structure-search runs were performed using both PFP and qSNAP, and low-cost candidate structures obtained from the two MLIPs were pooled before the subsequent DFT relaxations.

To quantify the agreement between experimental and simulated powder XRD patterns, we adopted a correlation-coefficient-based penalty,
\begin{align}
D &= 1 - \rho(I_{\mathrm{exp}}, I_{\mathrm{calc}}), \nonumber \\
\rho &=
\frac{\sum_k
\tilde{I}_{\mathrm{exp}}(2\theta_k)\,
\tilde{I}_{\mathrm{calc}}(2\theta_k)}
{\sqrt{\sum_k \tilde{I}_{\mathrm{exp}}^{\,2}(2\theta_k)}
\sqrt{\sum_k \tilde{I}_{\mathrm{calc}}^{\,2}(2\theta_k)}},
\end{align}
where $2\theta_k$ denotes grid points in the interval $[2\theta_{\min},2\theta_{\max}]$, and $\tilde{I}$ denotes an intensity profile shifted to zero mean.
Because $D=1-\rho$, this measure is scale-invariant and is sensitive to both peak positions and relative intensities.
The simulated intensity $I_{\mathrm{calc}}$ includes the Lorentz--polarization factor, and peak broadening is modeled by Gaussian smearing with a width of $\sigma=\SI{0.1}{\degree}$.
We examined two values of the weighting parameter, $\alpha=\SI{5}{\electronvolt}$/atom and \SI{10}{\electronvolt}/atom.

The cost function $F$ was minimized by simulated annealing (SA)~\cite{doi:10.1126/science.220.4598.671, WOODCOCK1971257} based on $NVT$ molecular dynamics, as implemented in \textsc{LAMMPS}~\cite{PLIMPTON19951, THOMPSON2022108171}.
The time step was \SI{1}{\femto\second}, and the temperature was linearly reduced from \SI{1000}{\kelvin} to \SI{0}{\kelvin} over total annealing times of 10--100\,\SI{}{\pico\second}.
To allow for long-period structural motifs, the initial configurations were prepared from $I4_1/a$-based supercells containing 8--128 atoms under periodic boundary conditions.

Because the cost function contains an XRD penalty term and the MLIP energy surface is only approximate, the structures obtained from the data-assimilation search are not guaranteed to correspond to minima on the DFT potential-energy surface.
All low-cost candidate structures were therefore further relaxed by variable-cell DFT calculations at the target pressures using \textsc{Quantum ESPRESSO}~\cite{Giannozzi_2009}, employing the PBE GGA functional and ultrasoft pseudopotentials~\cite{PhysRevB.41.7892}.
Plane-wave cutoffs of 40~Ry for the wave functions and 320~Ry for the charge density were used.
For eight-atom cells, a $16\times16\times16$ $k$-point mesh was sufficient to converge relative enthalpies within \SI{1}{\milli\electronvolt}/atom.
The resulting space groups were identified using \textsc{FINDSYM}~\cite{Stokes:zm5027}.
For the most promising low-enthalpy structures, relative enthalpies were recalculated using both the PBE GGA and r$^{2}$SCAN meta-GGA functionals as implemented in \textsc{VASP}.
In these calculations, we used a plane-wave cutoff of \SI{500}{\electronvolt} and a $16\times16\times16$ $k$-point mesh for eight-atom cells.

To examine lattice dynamical stability and superconducting properties, phonon dispersions and electron--phonon coupling coefficients were computed using density-functional perturbation theory (DFPT) in \textsc{Quantum ESPRESSO}.
The superconducting transition temperature $T_{\mathrm{c}}$ was estimated using the McMillan--Allen--Dynes formula~\cite{PhysRev.167.331,DYNES1972615,PhysRevB.12.905},
\begin{equation}
T_{\mathrm{c}}
= \frac{f_1f_2\omega_{\log}}{1.2}
\exp\left[
-\frac{1.04(1+\lambda)}
{\lambda - \mu^{\ast}(1+0.62\lambda)}
\right],
\end{equation}
where $\lambda$ is the electron--phonon coupling constant, $\omega_{\log}$ is the logarithmic average phonon frequency, and $\mu^{\ast}$ is the Coulomb pseudopotential.
The strong-coupling and shape-correction factors are
\begin{align}
f_{1} &=
\left[
1 + \left(
\frac{\lambda}{2.46(1+3.8\mu^{\ast})}
\right)^{3/2}
\right]^{1/3},
\\
f_{2} &= 1 +
\frac{
\left(\omega_2/\omega_{\log} - 1\right)\lambda^{2}
}{
\lambda^{2} + [1.82(1 + 6.3\mu^{\ast})(\omega_2/\omega_{\log})]^2
},
\end{align}
where
\begin{equation}
\omega_2
= \left(
\frac{2}{\lambda}
\int_0^\infty
\alpha^{2}F(\omega)\omega\, d\omega
\right)^{1/2}.
\end{equation}

The electron--phonon coupling constant and logarithmic phonon frequency were obtained from the Eliashberg spectral function $\alpha^{2}F(\omega)$,
\begin{align}
\lambda &=
2 \int_0^\infty
\frac{\alpha^{2}F(\omega)}{\omega}\, d\omega,
\\
\omega_{\log} &=
\exp\left[
\frac{2}{\lambda}
\int_0^\infty
\frac{\alpha^{2}F(\omega)}{\omega}
\log\omega \, d\omega
\right].
\end{align}

Phonon dispersions and electron--phonon coupling coefficients were computed using $16\times16\times16$ and $8\times8\times8$ $k$-point meshes for the dense and coarse electronic integrations, respectively, together with a $4\times4\times4$ $q$-point mesh.
We used $\mu^{\ast}=0.13$, which is a typical value for simple and transition-metal superconductors.

\subsection{Rietveld refinement}

Rietveld refinements of the experimental powder XRD profile at \SI{60}{\giga\pascal} were performed for the DFT-relaxed $R\overline{3}m$, $I4_1/a$, $Ibam$, and $R\overline{3}$ structural models using the \textsc{GSAS-II} software package~\cite{Toby:aj5212}.
For each model, least-squares fitting was carried out under the corresponding space-group symmetry constraints, with the scale factor, background, lattice parameters, profile parameters, and atomic coordinates refined.
The quality of the fit was assessed using the weighted-profile residual $wR$ and the goodness-of-fit (GOF).
\section{Results and Discussion}
\begin{figure*}
\includegraphics[width=\textwidth]{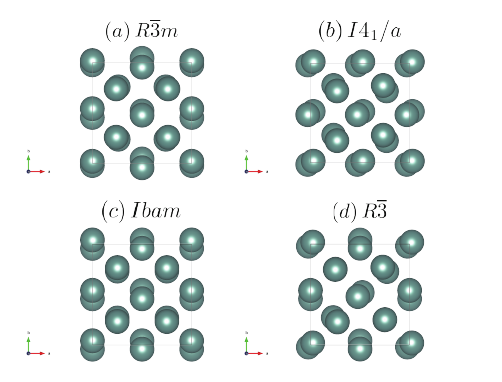}
\caption{
Crystal structures relevant to the dfcc phase of yttrium:
the three low-enthalpy phases obtained from the DA$+$DFT procedure,
(b) $I4_1/a$,
(c) $Ibam$, and
(d) $R\overline{3}$,
together with the previously proposed $R\overline{3}m$ model shown in (a) for comparison.
All models are shown in their DFT-relaxed geometries and visualized using \textsc{VESTA}~\cite{Momma:db5098}.
}
\label{fig:structures_dfcc}
\end{figure*}

The structures obtained directly from the data-assimilation (DA) calculations were not necessarily located at local minima on the DFT potential-energy surface (examples are shown in the Supplemental Material~\cite{SM}).
This is attributable to both the finite accuracy of the MLIPs used during the exploration stage and the propagation of uncertainties in the experimental diffraction profile $I_{\mathrm{exp}}$ into the penalty function $D$.
To obtain physically meaningful candidate structures, all DA-derived configurations were subsequently relaxed using DFT.

The DFT relaxations of the DA-derived candidates yielded three low-enthalpy phases: the previously reported $I4_1/a$ structure and two additional structures, $Ibam$ and $R\overline{3}$.
No DA-derived structure relaxed to the previously proposed $R\overline{3}m$ phase.
This absence of $R\overline{3}m$ among the relaxed minima is consistent with its dynamical instability, as will be discussed below from the phonon calculations.
The relaxed structures of the three DA-derived phases are shown in \cref{fig:structures_dfcc}, together with the previously proposed $R\overline{3}m$ model for comparison.
The $Ibam$ and $R\overline{3}m$ structures both retain a mirror plane parallel to the $bc$ plane, although their atomic displacements from the ideal fcc positions differ slightly.
In the cell representation adopted here, this difference is reflected in the lattice angle $\alpha$, which is $\SI{90.8}{\degree}$ for $R\overline{3}m$ and exactly $\SI{90}{\degree}$ for $Ibam$.
By contrast, the $R\overline{3}$ structure exhibits displacement patterns on the $(-110)$ plane similar to those of the $I4_1/a$ structure, indicating the loss of the mirror symmetry present in $R\overline{3}m$.
This is again reflected in the lattice angle, which becomes $\alpha = \SI{89.2}{\degree}$ in the adopted setting.
For completeness, the previously proposed $R\overline{3}m$ structure was also included in the subsequent enthalpy, phonon, and diffraction comparisons.
Notably, although the structural search was carried out using supercells containing up to 128 atoms, all three competitive phases can be reduced to primitive 8-atom cells.
This suggests that the essential structural motif of the dfcc phase can be represented within a compact primitive cell.

\begin{figure*}
\includegraphics{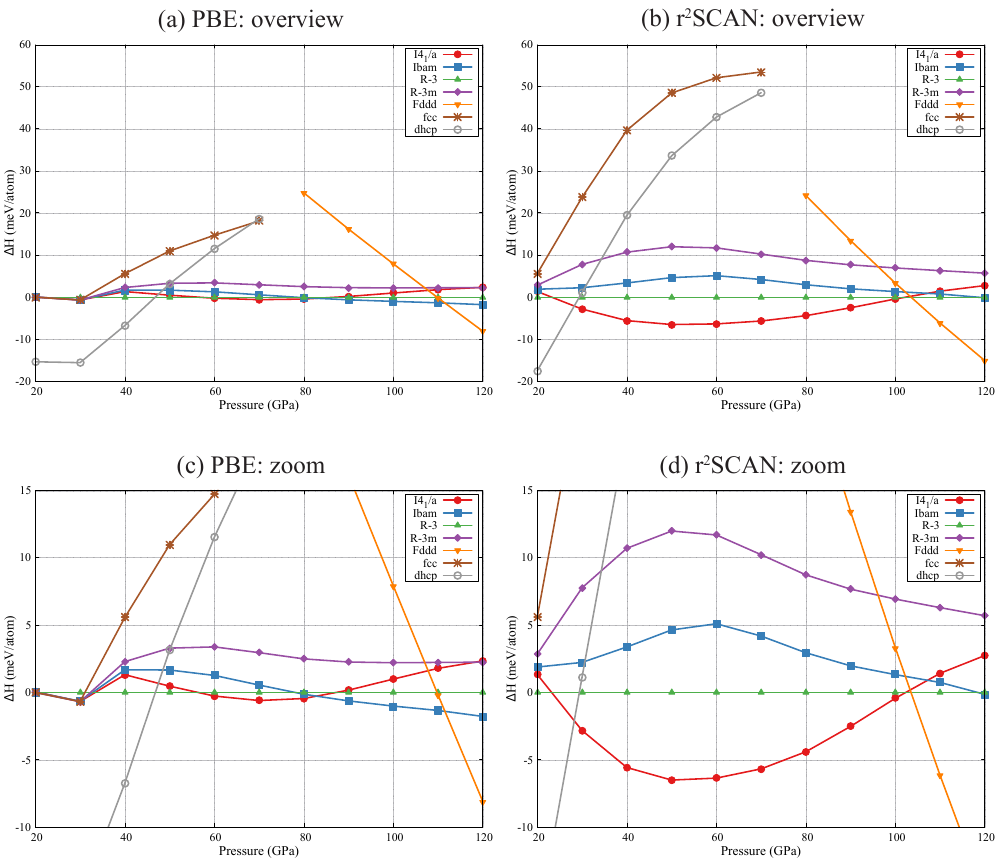}
\caption{
Pressure dependence of the relative enthalpies per atom of the four dfcc candidate phases, evaluated with respect to the $R\overline{3}$ phase, together with those of the dhcp, fcc, and $Fddd$ phases for reference.
(a) PBE GGA: overview.
(b) r$^{2}$SCAN meta-GGA: overview.
(c) PBE GGA: enlarged view of the low-enthalpy region.
(d) r$^{2}$SCAN meta-GGA: enlarged view of the low-enthalpy region.
The upper panels show the full enthalpy range including the dhcp, fcc, and $Fddd$ phases, whereas the lower panels highlight the small enthalpy differences among the dfcc candidate structures.
While PBE predicts near-degeneracy among $I4_1/a$, $Ibam$, and $R\overline{3}$, the r$^{2}$SCAN calculations yield a clearer energetic ordering, consistently stabilizing the $I4_1/a$ phase over the dfcc pressure range examined.
}
\label{fig:enthalpy_pbe_r2scan}
\end{figure*}

\Cref{fig:enthalpy_pbe_r2scan} shows the pressure dependence of the relative enthalpies of the four candidate phases.
For reference, the dhcp, fcc, and $Fddd$ phases are also included in the figure to place the dfcc candidates within the broader pressure-induced structural sequence of \ce{Y}, although the discussion below focuses on the dfcc-related structures.
As shown in Figs.~\ref{fig:enthalpy_pbe_r2scan}(a) and (c), PBE GGA predicts the $R\overline{3}m$ phase to be slightly higher in enthalpy than the other three, whereas $I4_1/a$, $Ibam$, and $R\overline{3}$ remain nearly degenerate.
In contrast, the r$^{2}$SCAN meta-GGA functional yields a clearer energetic ordering [Figs.~\ref{fig:enthalpy_pbe_r2scan}(b) and (d)]: the $I4_1/a$ phase is the most stable throughout the dfcc pressure range, followed by $R\overline{3}$, $Ibam$, and then $R\overline{3}m$.
Nevertheless, the relative enthalpies of the four dfcc candidates remain within about $\SI{10}{\milli\electronvolt\peratom}$, suggesting that more than one phase may be thermally accessible at room temperature.

\begin{figure*}
\centering
\includegraphics{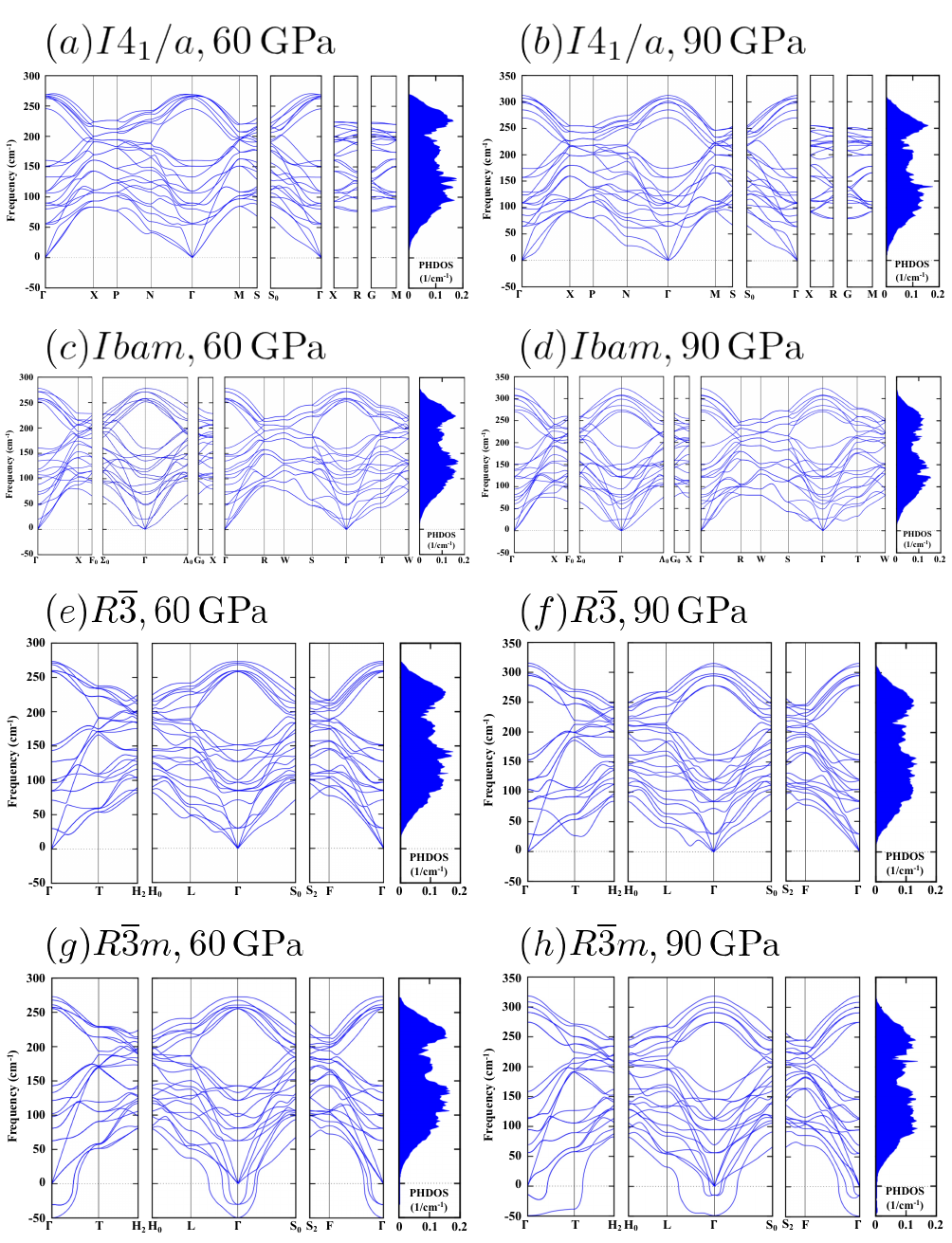}
\caption{
Phonon dispersions and phonon density of states (PHDOS) of the four dfcc candidate structures of \ce{Y} at \SI{60}{\giga\pascal} and \SI{90}{\giga\pascal}, computed using the PBE GGA functional.
(a),(b) $I4_1/a$;
(c),(d) $Ibam$;
(e),(f) $R\overline{3}$;
and (g),(h) $R\overline{3}m$ at \SI{60}{\giga\pascal} and \SI{90}{\giga\pascal}, respectively.
The PHDOS is shown on the right side of each panel.
No imaginary modes are found for $I4_1/a$, $Ibam$, and $R\overline{3}$ at either pressure, confirming their dynamical stability.
By contrast, $R\overline{3}m$ exhibits imaginary branches around the $\Gamma$ point at both pressures, indicating dynamical instability.
With increasing pressure, phonon branches in $I4_1/a$ soften along the $\Gamma$--$X$ and $\Gamma$--$N$ directions, whereas the soft branch in $Ibam$ hardens along the $\Gamma$--$\Sigma_0$ direction.
}
\label{fig:phonon_combined}
\end{figure*}

Phonon dispersions computed using the PBE GGA functional [\cref{fig:phonon_combined}] show no imaginary modes throughout the Brillouin zone for the $I4_1/a$, $Ibam$, and $R\overline{3}$ structures at pressures from 60--90\,\SI{}{\giga\pascal}, confirming their dynamical stability.
This result indicates that multiple distorted variants of the fcc lattice can exist as locally stable structures in the dfcc regime of \ce{Y}.
In the $I4_1/a$ structure, phonon branches along the $\Gamma$--$X$ and $\Gamma$--$N$ directions progressively soften under compression, whereas in the $Ibam$ structure the soft branch along the $\Gamma$--$\Sigma_0$ direction exhibits clear hardening.
These observations suggest that the $I4_1/a$ structure is favored over a wide pressure range, while the $Ibam$ structure becomes relatively more stable with increasing pressure.
Although the small enthalpy differences preclude a definitive conclusion, these dynamical trends are consistent with a possible pressure-induced transition from $I4_1/a$ to $Ibam$. By contrast, the $R\overline{3}m$ structure exhibits imaginary modes at and near the $\Gamma$ point, indicating dynamical instability.
Distorting the structure along the eigenvector of the unstable $\Gamma$-point mode and subsequently relaxing it leads to the $R\overline{3}$ structure, thereby ruling out $R\overline{3}m$ as a viable candidate for the dfcc phase.

\begin{figure*}
\includegraphics{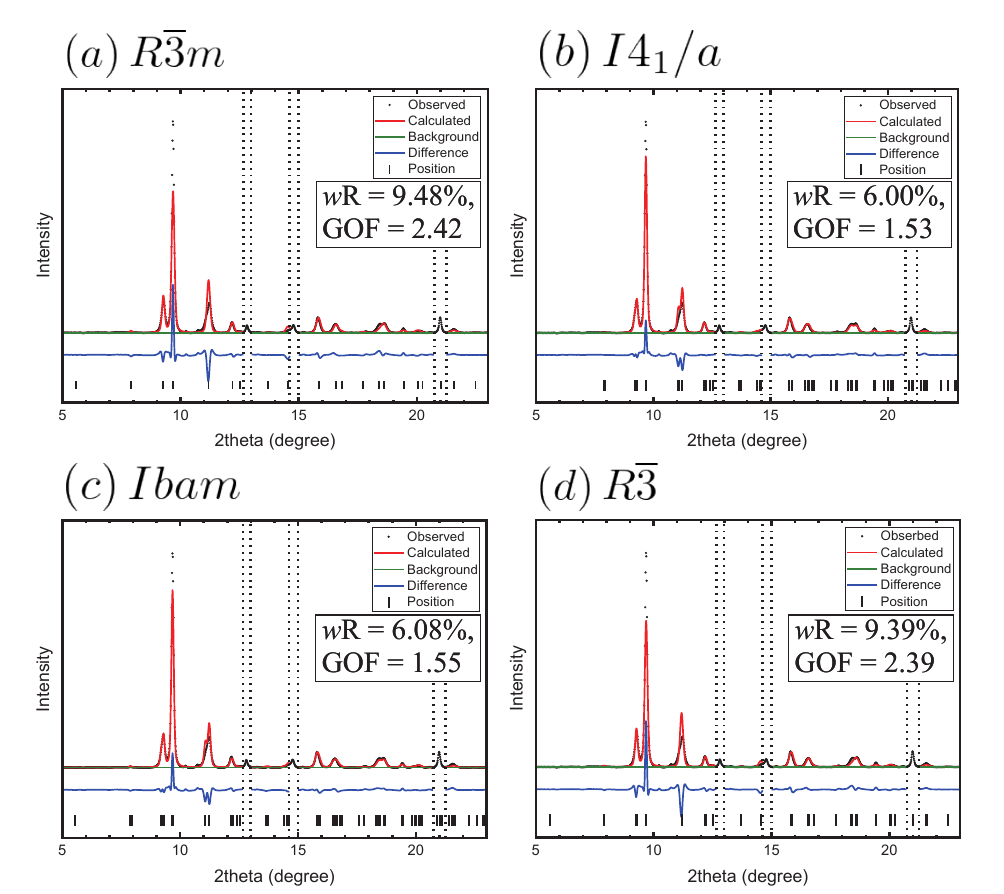}
\caption{
Rietveld refinement results for the four candidate dfcc structures at \SI{60}{\giga\pascal}.
(a) $R\overline{3}m$, (b) $I4_1/a$, (c) $Ibam$, and (d) $R\overline{3}$ models.
Experimental profiles (black circles), calculated profiles (red lines), background (green lines), difference curves (blue lines), and Bragg peak positions (black ticks) are shown.
The $2\theta$ regions indicated by the vertical dotted lines were excluded from the refinement because they contain diffraction peaks from \ce{Ne} used as the pressure-transmitting medium.
The $I4_1/a$ and $Ibam$ models yield better fits, with $wR = 6.00\%$ (GOF = 1.53) and $6.08\%$ (GOF = 1.55), respectively, than the $R\overline{3}m$ and $R\overline{3}$ models [$wR = 9.48\%$ (GOF = 2.42) and $wR = 9.39\%$ (GOF = 2.39)].
}
\label{fig:rietveld_60GPa}
\end{figure*}

To further assess the validity of the candidate structures, Rietveld refinements were performed for each structural model at \SI{60}{\giga\pascal}, and the agreement with the experimental powder XRD profiles was quantitatively evaluated.
The results are shown in \cref{fig:rietveld_60GPa}.
The $I4_1/a$ and $Ibam$ models yield $wR = 6.00\%$ and $6.08\%$, and GOF = 1.53 and 1.55, respectively, demonstrating better fits than the $R\overline{3}m$ and $R\overline{3}$ models, which give $wR = 9.48\%$ and $9.39\%$, and GOF = 2.42 and 2.39.
Notably, the residual factors and GOF values for $I4_1/a$ and $Ibam$ are nearly identical, indicating that these two models cannot be uniquely distinguished on the basis of Rietveld refinement alone.

A closer inspection of the fitted profiles reveals that the reproducibility of characteristic peak shapes in the low-angle region also depends on the structural model.
In particular, a doublet-like feature is observed in the experimental profile around $2\theta = \SI{11.2}{\degree}$, and this feature is reproduced more satisfactorily by the $I4_1/a$ and $Ibam$ models than by the other two structures.
These results indicate that, from the viewpoint of powder XRD profile fitting, $I4_1/a$ and $Ibam$ are the two most plausible structural models for the experimentally observed dfcc phase.

By contrast, the $R\overline{3}$ and $R\overline{3}m$ models give very similar residual factors and GOF values, indicating only minor differences in their ability to reproduce the experimental profiles.
This reflects the close similarity of their diffraction features.
Because structural assignments of dfcc phases in other rare-earth elements have often relied primarily on powder diffraction data, the present results suggest that the possibility of an $R\overline{3}$ phase should be reconsidered alongside the conventionally assigned $R\overline{3}m$ structure.

\begin{figure}
\includegraphics{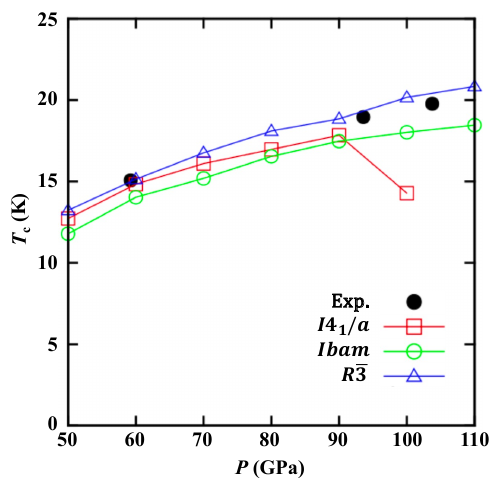}
\caption{
Pressure dependence of the superconducting transition temperature $T_{\mathrm{c}}$ for the dynamically stable dfcc candidates of \ce{Y}, calculated within the PBE GGA for the $I4_1/a$, $Ibam$, and $R\overline{3}$ structures, together with the available experimental data.
All three candidates reproduce the overall experimental trend within a few kelvin over much of the dfcc pressure range.
In particular, the calculated $T_{\mathrm{c}}$ values remain close to one another in the intermediate pressure range, whereas the $I4_1/a$ structure exhibits a marked decrease in $T_{\mathrm{c}}$ near \SI{100}{\giga\pascal}.
The $R\overline{3}m$ structure is not shown because it is dynamically unstable.
}
\label{fig:Tc}
\end{figure}

As a complementary, albeit indirect, probe of the candidate structures, superconducting transition temperatures ($T_{\mathrm{c}}$) were evaluated within the PBE GGA over the pressure range of 50--110\,\SI{}{\giga\pascal} [\cref{fig:Tc}].
Because the $R\overline{3}m$ structure is dynamically unstable, the comparison below focuses on the three low-enthalpy candidates, $I4_1/a$, $Ibam$, and $R\overline{3}$, noting that $I4_1/a$ becomes dynamically unstable at \SI{110}{\giga\pascal}.
The available experimental data indicate an increase in $T_{\mathrm{c}}$ from \SI{15.1}{\kelvin} to \SI{19.8}{\kelvin} with increasing pressure.
The calculated $T_{\mathrm{c}}$ values for $I4_1/a$, $Ibam$, and $R\overline{3}$ reproduce these experimental values within a few kelvin.
In particular, in the intermediate pressure range of 60--90\,\SI{}{\giga\pascal}, the differences among the three calculated $T_{\mathrm{c}}$ values remain within \SI{2}{\kelvin} and closely follow the interpolated experimental trend, indicating that superconducting behavior alone does not provide a decisive criterion for distinguishing among these structures.
At higher pressures, however, $T_{\mathrm{c}}$ for $I4_1/a$ decreases markedly at \SI{100}{\giga\pascal}, leading to a clear deviation from experiment.
Moreover, the $I4_1/a$ structure becomes dynamically unstable at \SI{110}{\giga\pascal}.
This behavior may reflect the increasing competitiveness of $Ibam$ or the onset of a transition to a higher-pressure phase such as $Fddd$.
Although the $R\overline{3}$ structure gives the closest overall agreement with the experimental $T_{\mathrm{c}}$, this result should be interpreted with caution in view of the energetic, dynamical, and diffraction-based considerations discussed above.
Taken together, the superconductivity results suggest that the observed experimental trend is more naturally explained by $I4_1/a$ and $Ibam$, possibly with pressure-induced structural changes or phase coexistence, rather than by a single structure over the entire dfcc pressure range.

\section{Conclusion}

In this work, we carried out a data-assimilation-based structure exploration of the dfcc phase of \ce{Y} by combining high-resolution powder XRD data with MLIPs.
By extending the search space to supercells containing up to 128 atoms, the DA$+$DFT procedure yielded three low-enthalpy phases: the previously reported $I4_1/a$ structure and two additional structures, $Ibam$ and $R\overline{3}$.
No DA-derived structure relaxed to the previously proposed $R\overline{3}m$ phase.
Although the search was performed over large supercells, all three DA-derived phases can be reduced to primitive 8-atom unit cells, indicating that the essential structural motif of the dfcc phase can be captured within a compact cell.

Enthalpy comparisons using both the PBE GGA and r$^{2}$SCAN meta-GGA functionals show that the previously proposed $R\overline{3}m$ phase is the least stable among the four structures considered.
While PBE predicts near-degeneracy among the $I4_1/a$, $Ibam$, and $R\overline{3}$ phases, r$^{2}$SCAN yields a clearer energetic ordering and consistently stabilizes the $I4_1/a$ phase throughout the dfcc pressure range.
Phonon calculations further show that $I4_1/a$, $Ibam$, and $R\overline{3}$ are dynamically stable over the pressure range examined, whereas $R\overline{3}m$ exhibits imaginary modes near the $\Gamma$ point and relaxes toward the $R\overline{3}$ structure.
These results explain why $R\overline{3}m$ was not obtained as a relaxed minimum in the DA$+$DFT search and rule it out as a viable candidate for the dfcc phase.
At the same time, the phonon trends indicate that $Ibam$ becomes increasingly competitive with respect to $I4_1/a$ toward the high-pressure side of the dfcc region.

Rietveld refinements at \SI{60}{\giga\pascal} further narrow the experimentally viable structural models to $I4_1/a$ and $Ibam$, both of which provide better fits to the powder XRD profiles than the $R\overline{3}m$ and $R\overline{3}$ models.
Because the refinement statistics for $I4_1/a$ and $Ibam$ are nearly identical, these two models cannot be uniquely distinguished by powder-diffraction fitting alone.
However, when the diffraction analysis is considered together with the r$^{2}$SCAN enthalpy ordering, $I4_1/a$ emerges as the most plausible structural model for the dfcc phase of \ce{Y}, with $Ibam$ remaining a close competing candidate, particularly at higher pressures.

As a complementary but indirect probe, the superconducting transition temperatures calculated for the dynamically stable candidate structures reproduce the experimental values within a few kelvin over much of the dfcc pressure range, indicating that superconducting behavior alone does not provide a decisive criterion for structural identification.
At the same time, the high-pressure evolution of $T_{\mathrm{c}}$ suggests that the experimentally observed behavior may be more naturally explained by $I4_1/a$ and $Ibam$, possibly with pressure-induced structural changes, phase coexistence, or the onset of a transition to the higher-pressure $Fddd$ phase.

Finally, the close similarity between the diffraction profiles of the $R\overline{3}$ and $R\overline{3}m$ structures highlights the difficulty of distinguishing between them using powder diffraction data alone.
This observation suggests that structural assignments of dfcc phases in other rare-earth elements may also warrant reexamination, and that the possibility of low-symmetry distortions with $R\overline{3}$ symmetry should be considered alongside the conventionally assigned $R\overline{3}m$ structure.

Overall, the present study demonstrates that data-assimilation-based structure exploration, combined with MLIPs, provides a powerful framework for identifying closely competing high-pressure structures that cannot be uniquely resolved from powder diffraction data alone.
This approach should be broadly applicable to other rare-earth systems and offers a general route toward accelerated structure determination under extreme conditions.
\begin{acknowledgements}
This work was supported by JSPS KAKENHI Grant Numbers JP25KJ1048, JP24K00544, and JP26K06996.
The authors thank the Supercomputer Center, the Institute for Solid State Physics, the University of Tokyo, for the use of the facilities.
Y.K. is supported by MEXT - Quantum Leap Flagship Program (MEXT Q-LEAP).
\end{acknowledgements}

\bibliography{apssamp}

\end{document}